# Hierarchical Magnetoelectric Coupling in Artificial Ferri-Order van der Waals Heterostructures

*Xilong Xu,*[1] *Li Yang**,[1,2]

[1]Department of Physics, Washington University in St. Louis, St. Louis, Missouri 63130, USA
[2]Institute of Materials Science and Engineering, Washington University in St. Louis, St. Louis, Missouri 63130, USA
E-mail: lyang@physics.wustl.edu

## Abstract

Multistate magnetoelectric coupling offers a promising route toward nonvolatile multibit memory, yet most ferroic systems remain limited to binary switching. Although van der Waals (vdW) heterostructures provide exceptional freedom to integrate distinct ferroic building blocks, a general and scalable strategy for engineering ordered multistate magnetoelectric responses remains lacking. Here, we propose artificial ferri-order vdW heterostructures as a general platform for hierarchical magnetoelectric coupling. Independently switchable interfacial magnetoelectric channels with different threshold energy barriers are sequentially activated under an applied electric field, generating an ordered progression of nonvolatile magnetic states. Using a $Sc_2CO_2$/bilayer-$CrGeTe_3$/$Nb_3Cl_8$ heterostructure as a prototype, first-principles calculations reveal four polarization-defined configurations mapped onto four distinct magnetic outputs. These states arise from two cooperative interfacial mechanisms: polarization reversal in $Nb_3Cl_8$ shifts inequivalent interfacial Cl atoms and tunes competing exchange pathways, whereas polarization switching in $Sc_2CO_2$ modifies the band alignment, drives charge transfer, and switches the magnetic order of $CrGeTe_3$. The four resulting states exhibit distinct anomalous Hall responses, enabling electrical readouts. These results establish artificial ferri-order engineering as a scalable platform for programmable multilevel ferroic architectures, nonvolatile multibit memory, multivalued logic, and reconfigurable spintronics.

## I. Introduction

Nonvolatile magnetoelectric coupling in multiferroic materials enables electric-field control of magnetic states without the large energy cost associated with current-driven spin switching, offering an attractive route toward low-power memory, logic, and neuromorphic computing. [1–4] However, most ferroelectric, magnetic, and magnetoelectric devices remain confined to a binary framework, in which two opposite polarization or magnetization states encode “0” and “1”. [5–7] While highly successful for conventional information technologies, this binary paradigm fundamentally limits information density and the realization of higher-order device functionalities. [8,9]

Ferri-order systems offer a natural pathway beyond this limitation. [10,11] In ferrielectrics and ferrimagnets, antiparallel but unequal electric dipoles or magnetic moments reside on nonequivalent sublattices, giving rise to intrinsic asymmetry, partial compensation, and multiple local degrees of freedom. [10–14] Coupling ferrielectric and ferrimagnetic orders could therefore transform multiple electrically addressable ferroic configurations into distinguishable magnetic outputs, enabling multibit memory, multivalued logic, and reconfigurable spintronic operations. [15–17] Realizing such multiferri functionality, however, remains challenging because materials possessing multiple controllable ferri-order degrees of freedom are rare, while magnetoelectric coupling in bulk systems is generally weak and largely unexplored in multiferri platforms. [3,10,18]

Van der Waals (vdW) heterostructures provide an ideal platform for realizing this concept because independently switchable ferroic building blocks can be vertically integrated while preserving their intrinsic properties. [19–22] More importantly, atomically sharp interfaces introduce multiple magnetoelectric coupling mechanisms, including exchange modulation, [23,24] charge transfer, [25–27] orbital hybridization, [28–30] and electrostatic band-alignment engineering, [31,32] that can be independently engineered. These characteristics enable artificial ferri-order architectures with programmable multistate magnetoelectric responses beyond naturally occurring materials.

Here, we propose an artificial ferri-order vdW heterostructure strategy for realizing electrically programmable hierarchical quaternary magnetoelectric coupling. By integrating two independently switchable polar layers with a bilayer $CrGeTe_3$ magnetic subsystem, we demonstrate four polarization-defined configurations that are mapped onto four distinct magnetic outputs through two microscopically different interfacial mechanisms. Because the corresponding interfacial magnetoelectric channels possess unequal switching barriers, they are activated sequentially under an applied electric field, producing an ordered evolution of nonvolatile magnetic states rather than a simple extension of binary switching. The four resulting states are electrically distinguishable through their anomalous Hall responses.

The remainder of this article is organized as follows: In Sec. II, we present the computational details of the simulation setup. The results of the hierarchical quaternary magnetoelectric coupling in the ferri-order van der Waals heterostructure are presented in Sec. III. Secs. IV and V discuss two interfacial mechanisms for this hierarchical coupling, respectively. Finally, we present the anomalous Hall response as the readout tools in Sec. VI.

## II. Computational Details

The first-principles simulations are performed using the Vienna ab initio Simulation Package (VASP). [33] The atomic structure optimization is performed within the generalized gradient approximation (GGA) and Perdew–Burke–Ernzerhof (PBE) exchange-correlation functional until the force on each atom is less than 0.02 eV/Å. [34,35] To better capture correlation effects and correct the underestimated band gaps in conventional DFT, the electronic structures and total energies relevant to magnetic-state comparisons were further evaluated using the HSE06 hybrid functional. [36] Due to the excessive complexity of the system, our calculations did not take spin-orbit coupling into account. The electronic iteration convergence criterion is set to 1 x $10^{-6}$ eV. To avoid spurious interactions between periodic images, a vacuum space of at least 20 Å is introduced. The k-mesh is set to 5 x 5 x 1 with a 550-eV cutoff energy. The DFT-D3 method is employed to include vdW interactions. [37] The FE transition pathway and energy barrier are calculated based on the nudged elastic band (NEB) approach. [38] The polarization is calculated by the modern polarization theory. [39,40]

## III. Hierarchical Magnetoelectric Coupling

**Figure 1** illustrates the conceptual design of hierarchical magnetoelectric coupling in artificial ferri-order vdW heterostructures. Owing to weak interlayer interactions and relaxed lattice-matching requirements, vdW materials can be vertically assembled from ferroelectric and magnetic building blocks with distinct polarization and magnetic degrees of freedom, thereby enabling artificial ferri-order architectures. The essential design principle, however, is not simply to generate multiple ferroic configurations, but to integrate multiple interfacial magnetoelectric channels with different microscopic origins, switching barriers, and dynamical responses. Because these channels respond to different electric-field thresholds and on different switching pathways, they can be activated sequentially rather than simultaneously. Each polarization configuration therefore selectively modifies a particular interfacial interaction, such as exchange coupling or electronic band alignment, and stabilizes a corresponding magnetic state. As schematically shown in **Figure 1**, this sequential activation converts the complexity of the artificial ferri-order architecture into an ordered progression of distinguishable magnetic outputs. The resulting staircase-like response defines hierarchical magnetoelectric coupling and provides multiple nonvolatile information states.

In the following, we demonstrate the design principle outlined above in an artificial ferri-order vdW heterostructure shown in **Figure 2(a)**, consisting of a ferroelectric $Sc_2CO_2$ monolayer, a bilayer $CrGeTe_3$ magnetic subsystem, and a multiferroic $Nb_3Cl_8$ monolayer. These constituents (< 0.5% lattice mismatch) are chosen from experimentally relevant and widely studied families of two-dimensional quantum materials. $CrGeTe_3$ is a representative layered vdW ferromagnetic semiconductor, with magnetism experimentally observed down to the few-layer and bilayer limits. [41,42] $Nb_3Cl_8$ is an experimentally available layered transition-metal halide that has attracted broad interest as a breathing-Kagome and correlated Mott material. [43–45] $Sc_2CO_2$, as a functionalized MXene monolayer, has been widely proposed as a prototype two-dimensional ferroelectric with robust out-of-plane polarization, [46] while MXenes themselves constitute a large and experimentally mature family of 2D materials. [47] The top $Nb_3Cl_8$ layer possesses both out-of-plane ferroelectric polarization and in-plane ferromagnetic order, while the bottom $Sc_2CO_2$ layer provides an independently switchable out-of-plane ferroelectric polarization. Together with the magnetic degrees of freedom of bilayer $CrGeTe_3$, these components form an artificial ferri-order architecture with multiple coupled electric and magnetic states.

In principle, these electric and magnetic degrees of freedom generate sixteen possible ferroic configurations, and we have performed the HSE06 hybrid-functional simulations to obtain their total energies, as summarized in **Table I**. Interestingly, the total-energy results reveal strong magnetoelectric coupling that lifts the energetic degeneracy among these states and selects a unique magnetic ground state for each polarization configuration. For example, when the polarization of $Sc_2CO_2$ points downward and that of $Nb_3Cl_8$ points upward, the $Nb_3Cl_8/CrGeTe_3$ interface favors ferromagnetic coupling, whereas the $CrGeTe_3$ bilayer favors antiferromagnetic alignment. Consequently, this AFM–FM state, with a total magnetic moment of $-1\ \mu_B$, becomes the magnetic ground state. As shown in **Table I**, for a fixed polarization configuration, most of the competing magnetic states are separated by several meV. This energy scale is sufficient to stabilize interlayer magnetic order in vdW magnets and is comparable to that underlying experimentally observed interlayer magnetic states, supporting robust polarization-selective control of the magnetic ground state. [23,24,48] Similar polarization-selective stabilization occurs for the remaining polarization states, producing four distinct magnetic ground states, denoted as A, B, C, and D in **Figure 2(a)**. The corresponding electric polarization, magnetic moments, and hysteresis loop are summarized in **Figure 2(b)**. These results demonstrate that four electrically programmable polarization configurations can be directly mapped onto four distinguishable magnetic outputs. The detailed electric and magnetic polarization information and stacking effects are provided in section I of the supplementary information. [49]

More importantly, the magnetoelectric coupling is hierarchical rather than merely a four-state correspondence determined by ground-state energetics. To investigate the switching pathways among different states, we have performed nudged elastic band (NEB) calculations for polarization reversal. The resulting energy landscape, shown in **Figure 2(b)**, reveals a sequence of unequal transition barriers that determines the evolution of the ferroic states under an external electric field. As a result, the system exhibits a staircase-like hysteresis loop with four discrete magnetic levels.

Starting from state A, where both $Nb_3Cl_8$ and $Sc_2CO_2$ polarizations point downward and the total magnetic moment is $+1\ \mu_B$, an upward electric field first drives the system into state B with a magnetic moment of $-1\ \mu_B$. As shown in **Figure 2(b)**, this transition is favored because the A→B barrier (~40 meV) is substantially lower than the competing A→C barrier (~120 meV). Physically, the A→B transition primarily involves polarization reversal of $Nb_3Cl_8$, whereas the A→C transition requires ferroelectric switching of $Sc_2CO_2$. The lower switching barrier of $Nb_3Cl_8$ therefore dictates the first step of the magnetoelectric response. Upon further increasing the electric field, the system subsequently evolves into state D with both polarizations upward and a magnetic moment of $13\ \mu_B$. When the electric field is reversed, the switching pathway is no longer symmetric: state D preferentially transforms into state C ($11\ \mu_B$) because the D→C barrier (~40 meV) is much smaller than the D→B barrier (~130 meV). Continued polarization reversal eventually returns the system to state A, completing the hysteresis cycle. The unequal switching barriers of the two ferroelectric layers lead to sequential rather than simultaneous polarization reversal, generating intermediate nonvolatile states with distinct magnetic outputs. Consequently, the total magnetic moment evolves through four well-separated plateaus, producing a hierarchical magnetoelectric response.

Our further analysis shows that the electrically programmable magnetic behavior does not originate from a single uniform interaction across the heterostructure. Instead, it is governed by two spatially separated yet cooperative interfacial magnetoelectric channels: the $Nb_3Cl_8$/CGT interface and the CGT/$Sc_2CO_2$ interface. The coexistence of these two distinct coupling pathways is central to the emergence of the hierarchical four-state response. To further elucidate the underlying mechanism, we next analyze these two interfacial coupling channels at the electronic-structure level.

## IV. $Nb_3Cl_8$/CGT interface

We first examine the microscopic origin of the magnetoelectric coupling at the $Nb_3Cl_8$/CGT interface, as illustrated in Figure 3(a), and find it is structure-controlled. In both $CrGeTe_3$ and $Nb_3Cl_8$, the transition-metal d orbitals are split by the crystal field. For Cr ions, the partially occupied ($\mathrm{t_{2g}}$) and ($\mathrm{e_g}$) orbitals give rise to the local magnetic moments. In $Nb_3Cl_8$, the Jahn–Teller distortion further lifts the degeneracy of the ($\mathrm{t_{2g}}$) manifold, producing a lower-energy ($\mathrm{t_{2g}^1}$) state and higher-energy ($\mathrm{t_{2g}^2}$) states. As a result, two competing interfacial exchange channels emerge across the $Nb_3Cl_8$/CGT interface.

The first channel, denoted as ($t_1$), is mediated by the interfacial Te and $Cl_1$ atoms and involves virtual hopping between the half-occupied Cr ($e_g$) orbitals and the half-occupied Nb ($t_{2g}^1$) state, as illustrated in **Figures 3(b)** and **3(c)**. This exchange pathway favors antiferromagnetic coupling. The second channel, denoted as ($t_2$), proceeds through the Te and $Cl_0$ atoms and involves hopping between the empty Nb ($t_{2g}^2$) states and the half-occupied Cr ($e_g$) orbitals, as shown in **Figures 3(b)** and **3(d)**. According to the Goodenough–Kanamori framework, this hopping process favors ferromagnetic coupling because the virtual excitation is stabilized by Hund's coupling on the Nb site. The observed magnetic ground state at the interface is therefore determined by the competition between the antiferromagnetic-favoring ($t_1$) channel and the ferromagnetic-favoring ($t_2$) channel.

A unique feature of Kagome $Nb_3Cl_8$ is the presence of two inequivalent interfacial Cl sublayers, labeled $Cl_0$ and $Cl_1$ in **Figure 3(b)**. These two Cl atoms occupy different vertical positions as a consequence of the ferroelectric distortion. Importantly, polarization reversal interchanges their relative heights: for the downward-polarized state (the case shown in **Figure 3(b)**), $Cl_1$ is located closer to the CGT layer than $Cl_0$, whereas the opposite arrangement is realized in the upward-polarized state.

Because interlayer exchange interactions in vdW heterostructures are highly sensitive to atomic separation, this polarization-driven structural reconstruction directly modifies the strengths of the competing exchange channels. When the polarization points downward (the case of **Figure 3(b)**), the shorter distance between $Cl_1$ and the CGT layer enhances the ($t_1$) hopping process, making the AFM exchange channel dominant. Upon polarization reversal, $Cl_0$ becomes the interfacial Cl atom closest to CGT, strengthening the competing ($t_2$) channel and promoting FM coupling. Consequently, ferroelectric switching of the $Nb_3Cl_8$ layer continuously tunes the balance between the FM- and AFM-favoring exchange interactions, providing a microscopic mechanism for electric-field control of the interfacial magnetic state. Notably, our previous study shows that this $Nb_3Cl_8$/CGT interfacial magnetoelectric coupling is robust against variations in high-symmetry interlayer stacking, [30] and this is also confirmed by our four-layer structure calculations presented in SI section I. [49]

## V. CGT/$Sc_2CO_2$ interface

We next examine the magnetoelectric coupling at the CGT/$Sc_2CO_2$ interface, as illustrated in Figure 4(a), and find it is electronic-structure controlled. In contrast to the $Nb_3Cl_8$/CGT interface, where polarization switching modifies competing exchange pathways through structural reconstruction, the CGT/$Sc_2CO_2$ interface operates through an electronic mechanism associated with polarization-controlled band alignment. Remarkably, switching the ferroelectric polarization of $Sc_2CO_2$ directly alters the magnetic ground state of the CGT bilayer. As seen from Table I and Figure 2(a), when the polarization points downward, the CGT bilayer retains its intrinsic AFM interlayer coupling. Upon reversing the polarization to the upward direction, the interlayer magnetic order of CGT bilayer is switched to an FM state. Notably, the interlayer magnetic coupling between $Nb_3Cl_8$ and CGT remains essentially unaffected by the switching of the $Sc_2CO_2$ polarization.

To understand the microscopic origin of this transition, we calculated the spin-resolved density of states (DOS) for the two polarization configurations, as plotted in **Figures 4(b1)** and **4(c1)**. A pronounced reconstruction of the electronic structure is observed upon polarization reversal. In the downward-polarization state, the heterostructure remains semiconducting with a finite band gap (~ 0.5 eV). In contrast, the upward-polarization state exhibits a finite DOS at the Fermi level, indicating the gap closing and the emergence of metallic behavior, as highlighted by the dashed circle in **Figure 4(c1)**.

The origin of this electronic transition is revealed by the band alignments shown in **Figures 4(b2)** and **4(c2)**. Reversing the out-of-plane polarization of $Sc_2CO_2$ changes the electrostatic potential across the interface and consequently shifts the relative band positions of $Sc_2CO_2$ and CGT. For the downward-polarization state, the band edges remain separated with a band gap around 0.5 eV, preserving the semiconducting character of the heterostructure. In contrast, the upward-polarization state drives the conduction and valence bands into contact, leading to band-gap closure and enhanced interfacial charge transfer. The resulting free carriers effectively dope the CGT bilayer. Consistent with previous theoretical studies, this carrier-induced doping favors the ferromagnetic interlayer coupling and ultimately stabilizes the FM ground state. [42,50]

Importantly, the band alignment and charge-transfer effects are evaluated using the HSE hybrid functional in this work. This treatment is essential because conventional DFT significantly underestimates the band gaps of the constituent materials, which can lead to substantial errors in predicting band offsets, charge transfer, and the resulting magnetic ground states. Because this mechanism originates primarily from the long-range electrostatic potential shift set up by the $Sc_2CO_2$ polarization, our calculations shows that the band-alignment-driven coupling to be comparatively less sensitive to interlayer stacking registry (see section I of the supplementary information [49]).

Together with the geometry-controlled exchange mechanism at the $Nb_3Cl_8$/CGT interface, this band-alignment-driven magnetic switching forms the second magnetoelectric channel in the heterostructure. The coexistence of these two microscopically distinct yet cooperative mechanisms ultimately gives rise to the hierarchical four-state magnetoelectric response observed in the artificial ferri-order vdW heterostructure.

## VI. Anomalous Hall conductivity readout:

To experimentally detect the hierarchical magnetoelectric coupling and explore its device potential, we propose using the anomalous Hall conductivity (AHC) as an electrical readout of the four polarization-defined ferroic states. As illustrated in Figure 5(a), a Hall-bar device geometry can be fabricated from the proposed heterostructure, where the ferroic state is controlled by an external out-of-plane gate electric field and subsequently detected through the transverse Hall response. Because the four polarization configurations correspond to four distinct magnetic ground states, polarization switching is expected to induce discrete changes in the Hall conductance, providing a direct electrical signature of the hierarchical magnetoelectric coupling.

Compared with local magnetic probes or structural characterization techniques, AHC offers a non-destructive and device-compatible readout mechanism. Once the four ferroic states are electrically written through polarization switching, they can be distinguished by their corresponding Hall responses without requiring direct imaging of the underlying magnetic or ferroelectric domains. Such an all-electrical operation scheme is particularly attractive for practical implementations of nonvolatile multibit memory and multivalued logic devices.

To quantitatively evaluate the transport response, we calculate the intrinsic anomalous Hall conductivity using the Kubo formula, [51,52]

$$\sigma_{xy} = -\frac{e^2}{\hbar}\sum_n \int_{\mathrm{BZ}} \frac{d^2k}{(2\pi)^2} f_{n\mathbf{k}} \Omega_n^z(\mathbf{k}) \quad , \quad (1)$$

where $f_{nk}$ is the Fermi–Dirac distribution function, and $\Omega_n^z$ denotes the Berry curvature of the $n$-th electronic band. The integration is performed over the entire Brillouin zone. The simulation details are provided in section II of the supplementary information. [49] Because the Berry curvature is highly sensitive to the electronic structure and magnetic configuration, the AHC provides an effective probe of the polarization-controlled magnetic states in the heterostructure.

The calculated AHC spectra are presented in **Figure 5(b)**. The four ferroic states (A, B, C, and D) exhibit clearly distinguishable Hall responses over a broad energy range, demonstrating that the polarization-controlled magnetic configurations can be uniquely identified through their transverse transport signatures. Notably, the differences remain evident around the Fermi level and become even more pronounced when the chemical potential is shifted. For example, at a chemical potential of approximately -0.2 eV, the A ($-9.1 \times 10^{-3}$ S/cm), B ($1.06 \times 10^{-3}$ S/cm), C ($-7.02 \times 10^{-3}$ S/cm), and D ($-2.1 \times 10^{-3}$ S/cm) states can be readily distinguished. This suggests that electrostatic gating or carrier tuning can further enhance the separation between different readout states. The one-to-one correspondence between ferroic configurations and Hall responses establishes a direct link between the internal magnetoelectric state and a measurable electrical output.

More broadly, this work establishes artificial ferri-order as a general design principle for programmable magnetoelectric materials. Rather than relying on the limited availability of naturally occurring multiferri compounds, independently switchable ferroic building blocks can be assembled into vdW heterostructures, where distinct interfacial magnetoelectric channels are engineered to possess different switching barriers and threshold fields. Their sequential activation under an applied electric field enables hierarchical magnetoelectric switching and an ordered progression through multiple nonvolatile ferroic states. Notably, the present heterostructure contains two independently switchable magnetoelectric channels and therefore supports four nonvolatile states. More generally, an architecture containing (N) independently switchable binary channels could, in principle, generate up to ($2^N$) polarization configurations and corresponding magnetic outputs, provided that the channels remain sufficiently decoupled and their switching thresholds are distinguishable. Because this strategy relies on modular interface engineering rather than material-specific chemistry, it can be extended to a broad range of two-dimensional ferroelectric, magnetic, and multiferroic materials. The ability to independently design, combine, and sequentially address multiple magnetoelectric channels thus provides a scalable route toward programmable multilevel ferroic architectures for multibit memory, multivalued logic, neuromorphic computing, and reconfigurable spintronic devices.

In summary, we have proposed an artificial ferri-order van der Waals heterostructure strategy to realize hierarchical multistate magnetoelectric coupling beyond conventional binary switching. By integrating a ferroelectric $Sc_2CO_2$ monolayer, a bilayer $CrGeTe_3$ magnetic subsystem, and a multiferroic $Nb_3Cl_8$ monolayer, the designed heterostructure supports four polarization-defined states that are selectively mapped onto four distinct magnetic ground states. Hybrid-functional first-principles calculations reveal that this four-state response is governed by two spatially separated yet cooperative interfacial mechanisms. At the $Nb_3Cl_8$/$CrGeTe_3$ interface, polarization reversal changes the relative positions of inequivalent Cl sublayers, thereby tuning competing FM- and AFM-favoring exchange pathways. At the $CrGeTe_3$/$Sc_2CO_2$ interface, polarization-controlled band alignment induces charge transfer into the $CrGeTe_3$ bilayer and switches its interlayer magnetic coupling. The unequal switching barriers of the two polar layers further lead to sequential polarization reversal, producing a staircase-like magnetoelectric response with four nonvolatile magnetic plateaus. In addition, the four ferroic states exhibit clearly distinguishable anomalous Hall conductivity spectra, providing a practical all-electrical readout scheme. These results establish artificial ferri-order vdW heterostructures as a flexible platform for nonvolatile multibit memory, multivalued logic, neuromorphic computing, and reconfigurable spintronic devices, and suggest broader opportunities for engineering higher-order magnetoelectric functionalities through rational vdW interface design and ferroic building-block integration.

**Author Informations**

Corresponding Author:

**Li Yang** - Department of Physics and Institute of Materials Science and Engineering, Washington University in St. Louis, St. Louis, Missouri 63130, USA

Email: lyang@physics.wustl.edu

Author:

**Xilong Xu** - Department of Physics, Washington University in St. Louis, St. Louis, Missouri 63130, USA

Email: xilong@wustl.edu

**Notes**

The authors declare no competing financial interest.

**Data Availability Statement**

There are no publicly available research data or software supporting this manuscript. Requests for further information or data should be sent to the authors

**Acknowledge:**

This work is supported by the Department of Energy (DOE), Office of Science, Basic Energy Sciences under Award No. DE-SC0026312. The simulation used Anvil at Purdue University through allocation DMR100005 from the Advanced Cyberinfrastructure Coordination Ecosystem: Services & Support (ACCESS) program, which is supported by National Science Foundation grants #2138259, #2138286, #2138307, #2137603, and #2138296.

**Table I.** Relative energies, in meV, of different magnetic configurations under four polarization states obtained from HSE calculations. The lowest-energy magnetic configuration for each polarization state is set to zero. In the polarization notation ($P_{down}/P_{up}$), the first and second entries denote the polarization directions of the $Sc_2CO_2$ and $Nb_3Cl_8$ layers, respectively. In the magnetic notation (AFM/FM), the first entry refers to the magnetic coupling within the bilayer CGT subsystem, whereas the second entry refers to the magnetic coupling between CGT and $Nb_3Cl_8$.

| | AFM-AFM ($1\mu_B$) | AFM-FM ($-1\mu_B$) | FM-AFM ($11\ \mu_B$) | FM-FM ($13\ \mu_B$) |
|---|---|---|---|---|
| $P_{down}$-$P_{down}$ | **0 (A)** | 0.43 | 2.64 | 3.09 |
| $P_{down}$-$P_{up}$ | 5.37 | **0 (B)** | 7.50 | 2.10 |
| $P_{up}$-$P_{down}$ | 1.06 | 1.32 | **0 (C)** | 0.28 |
| $P_{up}$-$P_{up}$ | 7.19 | 7.18 | 6.07 | **0 (D)** |

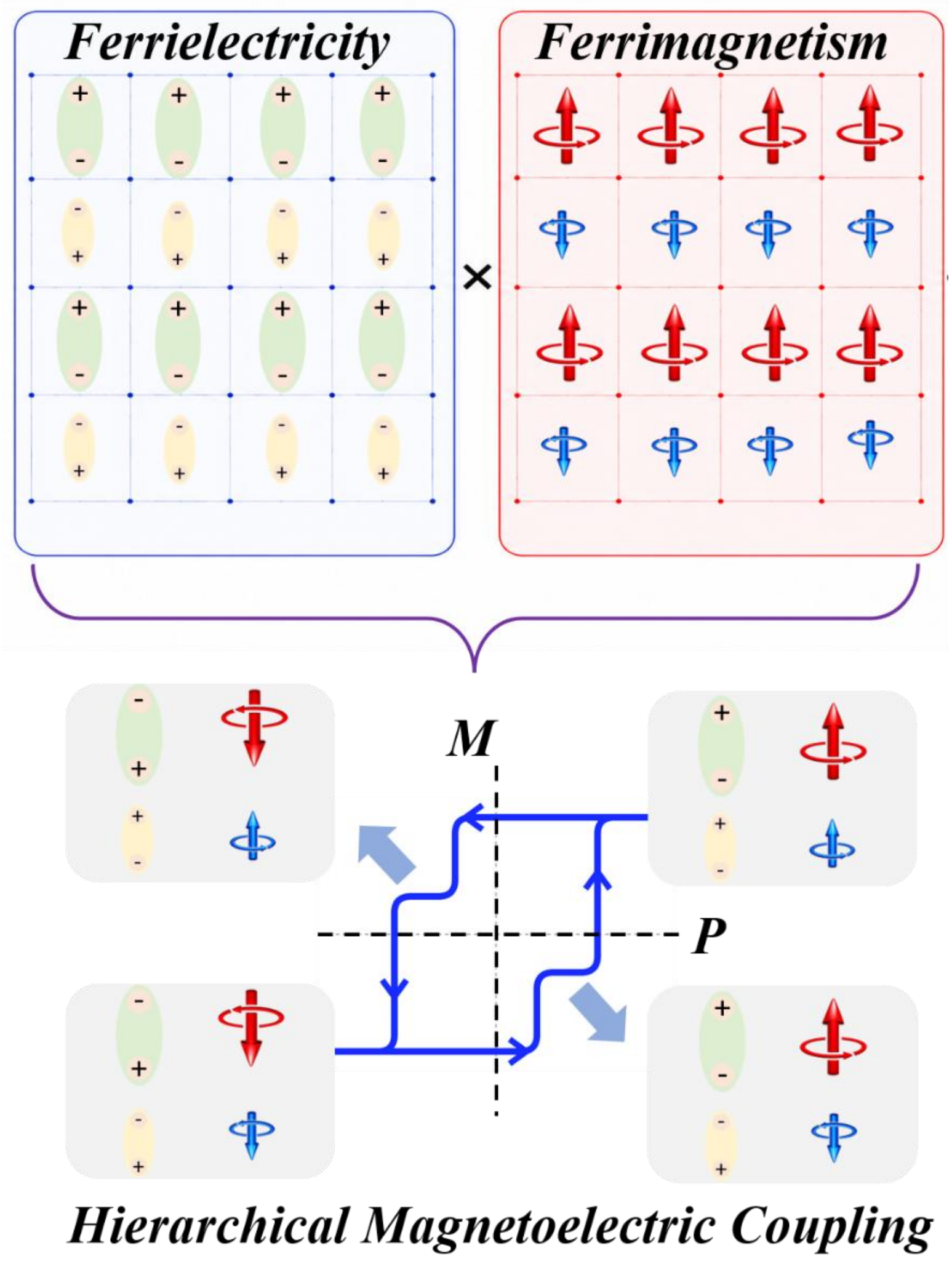

**Figure 1.** Conceptual design of ferri-order-enabled multistate magnetoelectric coupling. Ferrielectric and ferrimagnetic orders, composed of antiparallel but unequal electric dipoles and magnetic moments, are coupled to generate four discrete output states, enabling quaternary-state memory functionality.

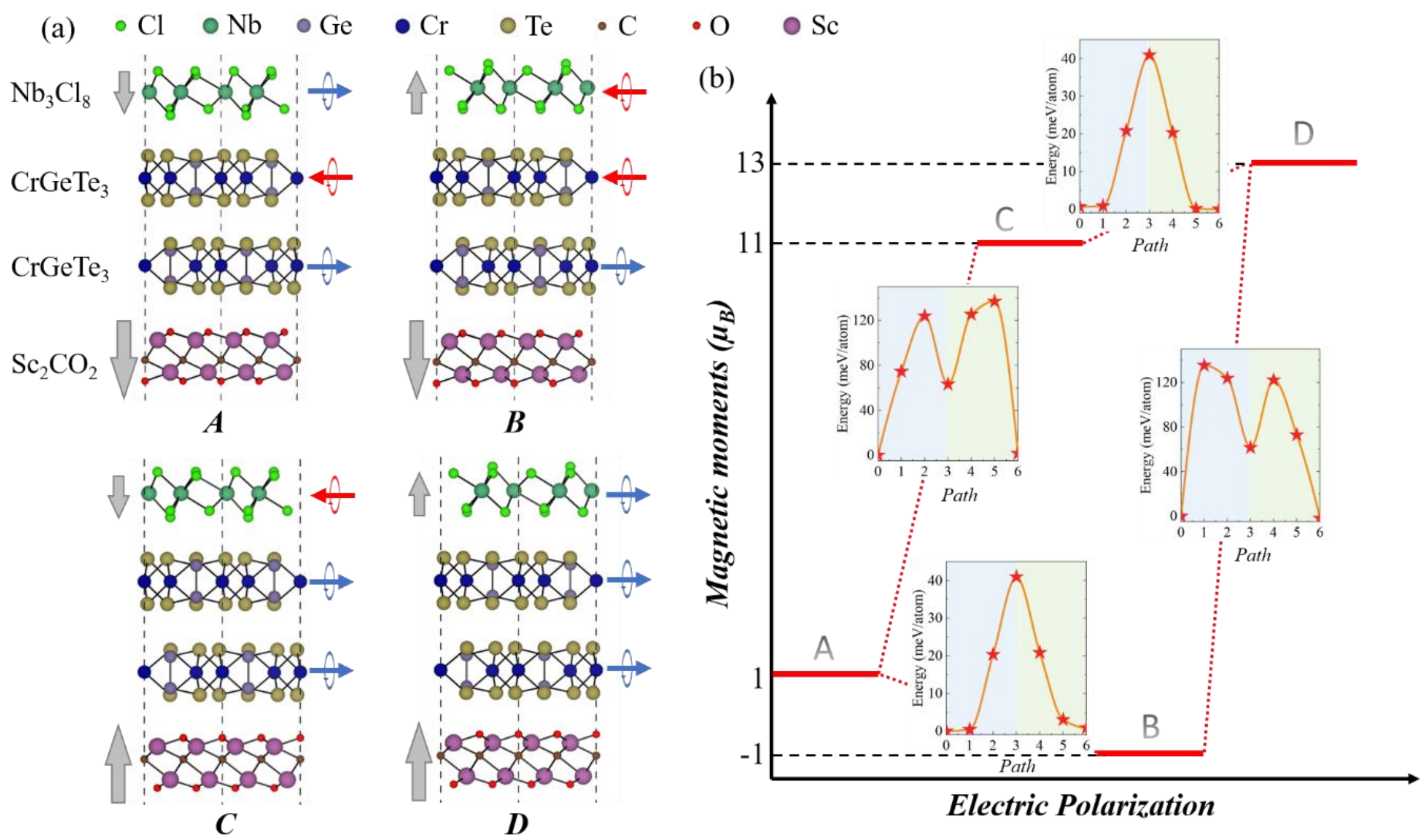

**Figure 2.** Multistate magnetoelectric behavior of the designed van der Waals heterostructure. (a) Four polarization-defined configurations and the corresponding magnetic states in $Nb_3Cl_8$/bilayer $CrGeTe_3$/$Sc_2CO_2$ heterostructure. (b) Schematic switching landscape connecting the four states A–D, showing the stepwise evolution of the magnetic output. The insets indicate the calculated transition barriers between neighboring polarization states.

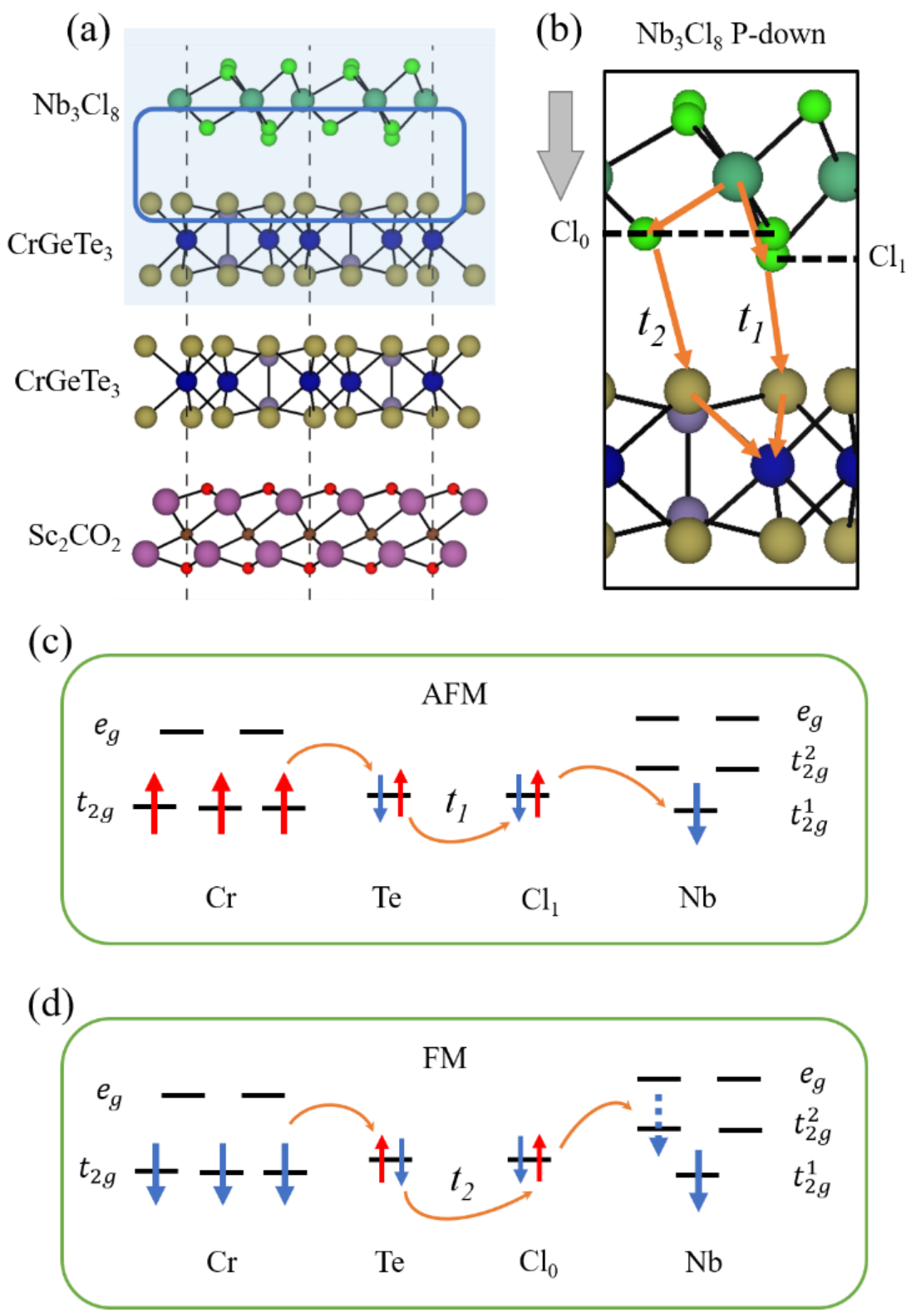

**Figure 3.** (a) The highlighted $CrGeTe_3/Nb_3Cl_8$ interface. (b) The two competing interlayer hopping paths, $t_1$ and $t_2$, mediated by inequivalent Cl sublayers. (c) Schematic of the AFM-favoring $t_1$ exchange channel involving virtual hopping between Cr and Nb states mediated by $Cl_1$ sublayers. (d) Schematic of the FM-favoring $t_2$ exchange channel involving virtual hopping between Cr and Nb states mediated by $Cl_0$ sublayers. Polarization reversal changes the relative interfacial heights of $Cl_0$ and $Cl_1$, thereby tuning the balance between these two exchange pathways.

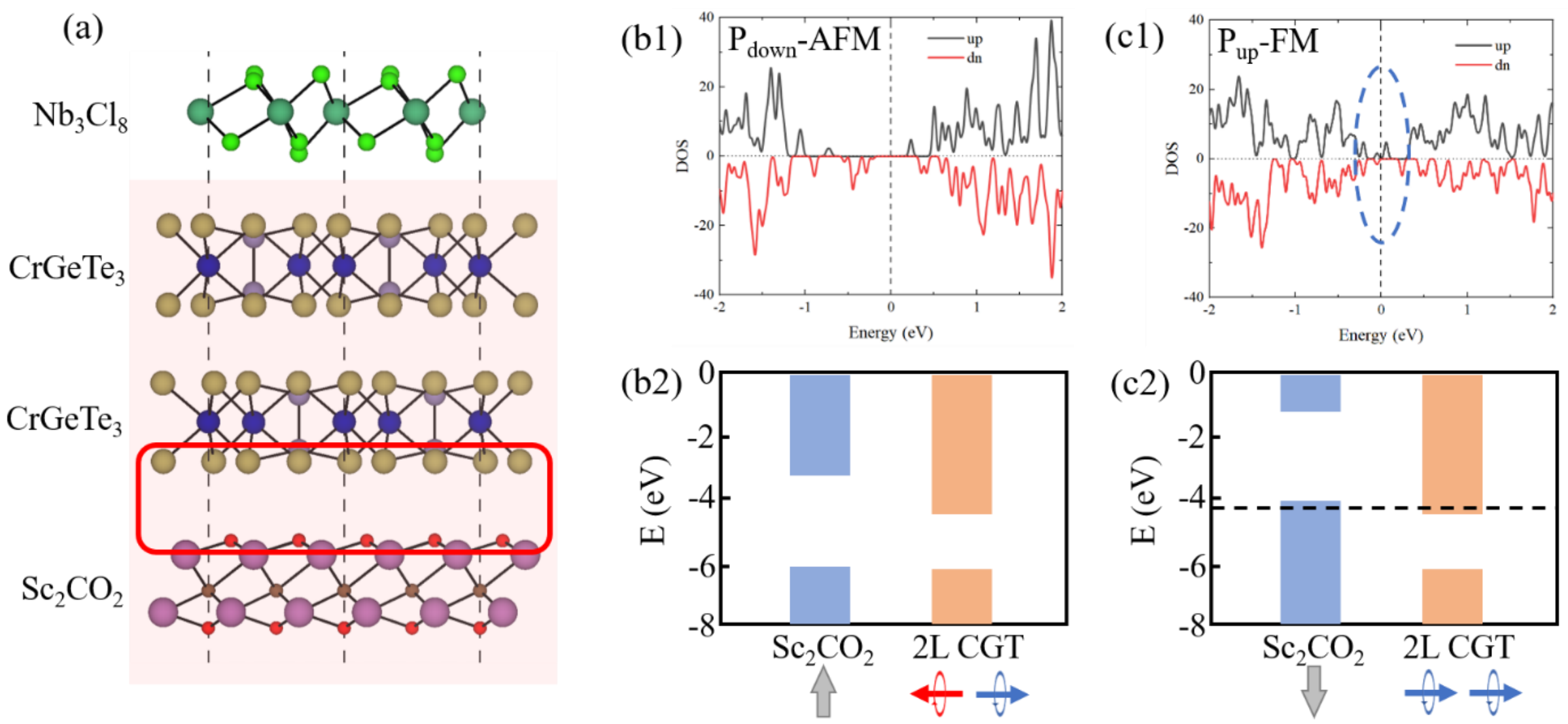

**Figure 4.** (a) The highlighted lower $CrGeTe_3/Sc_2CO_2$ interface. (b1,c1) Spin-resolved DOS for the $P_{down}$-AFM and $P_{up}$-FM states, respectively. The $P_{down}$ state preserves a semiconducting AFM configuration, whereas the $P_{up}$ state induces band overlap near the Fermi level, leading to metallic behavior and FM coupling. (b2,c2) Corresponding band-alignment diagrams showing the polarization-driven band shift, charge transfer, and carrier doping responsible for the AFM-to-FM transition.

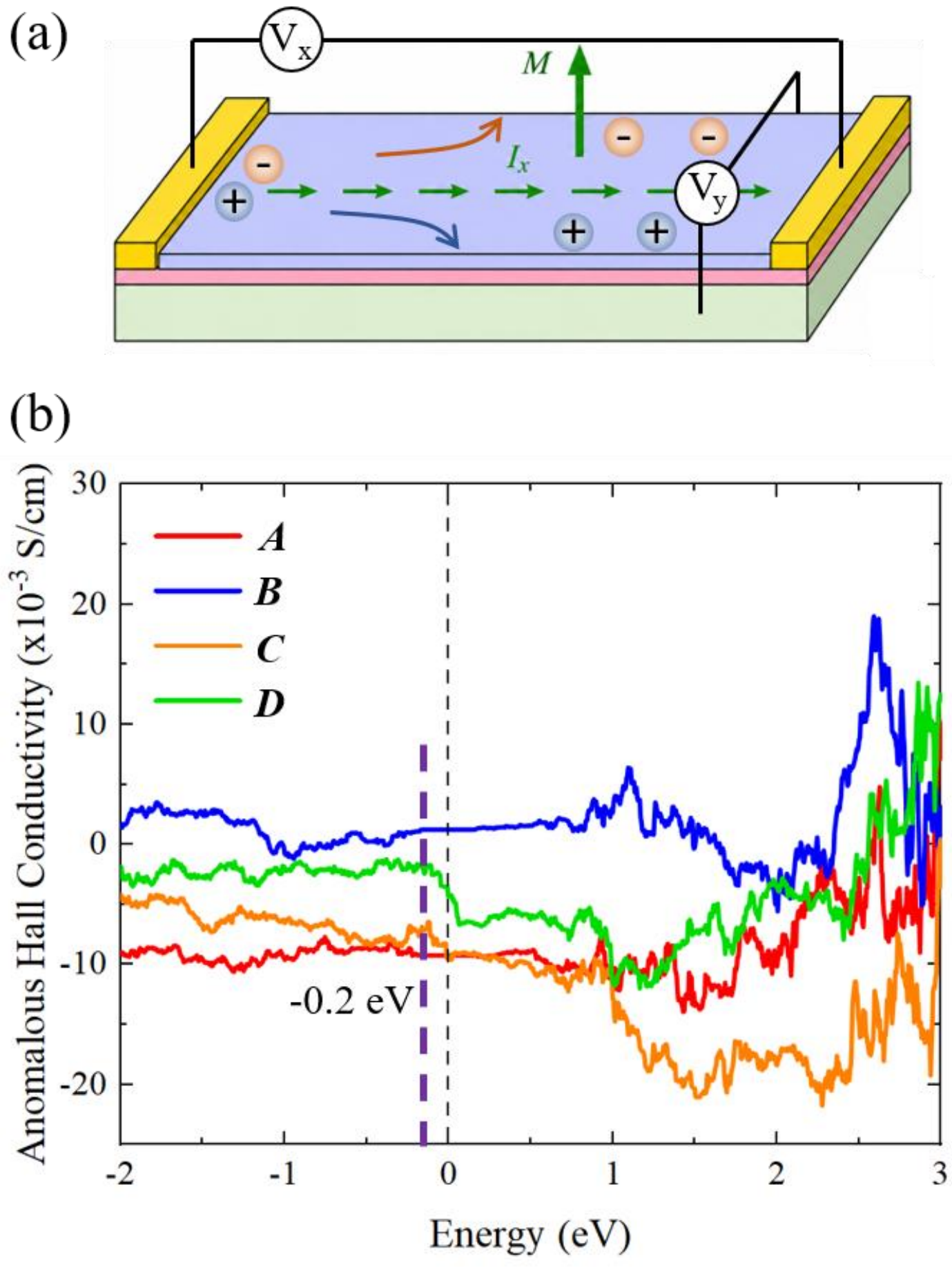

**Figure 5.** Anomalous Hall conductivity of the four polarization-defined states. (a) Schematic Hall-bar device for electrical readout. (b) The calculated anomalous Hall conductivity for the four different polarization states, showing distinguishable transport responses that can be used to identify the different magnetoelectric states.